\documentstyle[aps,preprint]{revtex}

\begin{document}

\draft

\title{Implications of quaternionic dark matter}

\author{S.\ P.\ Brumby\footnote{spb@physics.unimelb.edu.au},
B.\ E.\ Hanlon and G.\ C.\ Joshi}

\address{Research Centre for High Energy Physics, \\
School of Physics, University of Melbourne, \\
Parkville, Victoria 3052, Australia}
\date{\today}
\maketitle
\begin{abstract}
Taking the complex nature of quantum mechanics which we observe today
as a low energy effect of a broken quaternionic theory we explore
the possibility that dark matter arises as a consequence of this
underlying quaternionic structure to our universe. We introduce a low
energy, effective, Lagrangian which incorporates the remnants of a local
quaternionic algebra, investigate the stellar production of the resultant
exotic bosons and explore the possible low energy consequences of our
remnant extended Hilbert space.
\end{abstract}
\pacs{{\it keywords:} quaternion dark matter paraphoton
{\it Report number:} UM--P--96/90; RCHEP 96/12.}

\section{Introduction}
Observational evidence and big bang nucleosynthesis bounds point to
the existence of non-baryonic dark matter.  The various dark matter
candidates which have been explored in the literature all derive from
attempts to extend the standard model either in purely phenomenological
ways, or more fundamentally by seeking to incorporate the standard model 
into larger structures. Of the various fundamental approaches, one
promising alternative which has not received much attention is a
quaternionic quantum field theory (QQFT).

In fact, there are many ways to produce a quaternionic generalisation
of the mathematical structure of quantum field theory.  The axiomatic
analysis of quaternionic quantum mechanics undertaken by Birkhoff and
von Neumann \cite{bvn} allows many formulations to be considered
\cite{fjss,hb,nj,beh,adl,bja}.
To constrain some of these possibilities we take the obvious success of
the complex formulation of high energy  physics as a clear sign that
nature is {\em locally} describable by complex quantum field theory.
By exploiting this local interpretation and the extended automorphism
group of the quaternions we can introduce a local SU(2) gauge
symmetry which is naturally bound to our underlying construction.
We do not propose to construct a full quantum field theory of quaternionic
states.  Rather, we consider the low energy effects of a quaternionic
theory which, by some mechanism, has broken to the local complex structure
we observe today.  We wish to investigate if the resulting exotic bosons
which arise as remnant degrees of freedom of the full SU(2) theory can be 
identified as dark matter candidates.  Such a notion would imply that
quaternionic quantum effects may arise at galactic and cosmological
scales.

The breaking is realised by singling out a particular quaternion which is
ultimately identified with the imaginary unit of complex quantum field
theory. In this way an exotic U(1) survives to low energies.  Our exotic
bosons are coupled to normal matter via non-renormalisable terms in
an effective Lagrangian. Our theoretical construct is sufficiently neat
that only a relatively small number of parameters are needed to describe
the new physics.

The crucial point is that we are seeking to encompass the introduction of
new dark matter candidates and extensions of current physics from 
a fundamental shift in our mathematical description of nature.  In this
way we differ from simply introducing a phenomenological SU(2),
from which broader issues, such as the nature of the quantum theory,
cannot be discussed.  
Indeed, this approach places new constraints on our effective Lagrangian
which are not prescribed by a purely phenomenological theory.

Our aims in this paper are twofold: to explore if this notion is
phenomenologically compatible with known astrophysical bounds and
to introduce possible consequences of our underlying extended quantum
mechanical structure. This will be addressed by first introducing
our effective lagrangian in
the next section, from which we calculate an astrophysical bound in the 
context of solar physics.  Next, we consider the question of
the cosmological significance of our exotic vector bosons.
Finally, we shall speculate on the fundamental Lagrangian, the dynamics
which underlies our effective theory and the possible phenomena which may
be encapsulated by it.

\section{Model}

Studies of QQFT have most often started by considering a Hilbert space
over the quaternionic field, for which the quantum mechanical amplitudes
must, in general, be quaternionic and hence do not commute.  We shall
instead introduce a simplified model which, we maintain, retains the
spirit of the original while holding out the best hope of connecting
with experimental physics.  

Our approach is to consider the complex nature we observe today as a low
energy remnant of a full quaternionic theory at high energies. We thus
first consider a space-time with a set
of scalar fields $i(x_\mu)$, $j(x_\mu)$, and $k(x_\mu)$, which locally
define a basis for the set of pure imaginary quaternions. Being local,
invariance is required under the extended set of automorphisms of the 
quaternionic algebra which vary continuously over space-time,
leading to an SU(2) like exotic sector.  We relate amplitudes
at different places
by considering local quaternionic phase transformations of the type
$\phi\rightarrow q\phi q^{-1}$, where $q$ is a pure imaginary quaternion
of unit magnitude. Note that at this stage all the fundamental fields are
presumed to be quaternionic. It is worth noting, however, that we do not
identify this sector with 
the fundamental SU(2) in electroweak theory, nor with any approximate
SU(2) in hadronic physics.  

The gauge-covariant derivative for this exotic gauge symmetry follows
the usual pattern of Yang-Mills
theory, but with a quaternionic valued potential,
$D_{\mu}\phi =\partial_{\mu}\phi +\frac{1}{2}(Q_{\mu}\phi -\phi Q_{\mu})$, where
\begin{equation}
Q_{\mu}(x) \equiv \bar{A}_{\mu}(x) i(x) + B_{\mu}(x) j(x)
+ B'_{\mu}(x) k(x)\,.
\end{equation}
Our choice of notation follows, {\em a posteriori}, from our 
interpretation of the phenomenological implications of our model.

{}From this we construct the field strength tensor,
$K_{\mu\nu} = D_{\mu}Q_{\nu} - D_{\nu}Q_{\mu}$
and exhibit the Lagrangian density for the new gauge sector,
\begin{equation}\label{Lfree}
{\cal L}_{\sc q} = \frac{1}{4}K^{\mu\nu}K_{\mu\nu}
+ \frac{\lambda}{2}|D_{\mu}i|^2 + \frac{\lambda'}{2}|D_{\mu}j|^2 \,.
\end{equation}

This Lagrangian differs from that of Finkelstein, {\em et
al.}\ \cite{fjss}
in that {\em two} pure imaginary quaternionic fields are singled out
and considered as fundamental fields. This requires comment.
Firstly, it is
important to notice that the $i$ and $j$ fields are dimensionless.
Scale is introduced by the constants $\lambda$, $\lambda'$, which are
independent in the most general case, and have the dimension of 
mass squared (${\cal L}_{\sc q}$ is composed of dimension 4 operators).
The components of the $Q$ potential have the usual dimension of
vector bosons.
Furthermore, we have not included a $(\lambda''/2)|D_{\mu}k|^2$, as at
each point
of space-time, a consistent definition of the quaternionic algebra
requires $k(x) \equiv i(x)\,j(x)$, so there are no additional dynamical
degrees of freedom. 

Quaternionic SU(2) transformations which leave the field $i(x)$ invariant
define a U(1) subgroup of the general quaternionic group. We presume that
via some mechanism this field has indeed been especially singled out. In
this way our fully quaternionic theory can be brought into contact with
the usual complex quantum theory.
In order to interpret the physical content of our model, we make a
particular local quaternionic gauge transformation, to what we call
the i-gauge (strongly analogous to the unitary gauge of electroweak
physics). There always exists a quaternionic gauge transformation,
$q_{\rm i}$,
taking $i(x_{\mu})$ to a particular pure imaginary unit quaternion $i$,
$q_{\rm i}\,i(x_{\mu})\,q_{\rm i}^{-1} = i$ \cite{fjss}.
We associate this
imaginary with the ordinary complex imaginary of complex quantum theory.
In this particular
gauge, the quaternionic potential is
$Q_{\mu}(x) = \bar{A}_{\mu}(x)\,i + [B_{\mu}(x)+ B'_{\mu}(x)\,i]j'(x)$,
where $j'(x)\!=\!q_{\rm i}\,j(x)\,q_{\rm i}^{-1}$ is defined up to
a complex phase,
\begin{equation}
j'(x) = \exp[\vartheta(x)\,i]\,j=\cos\vartheta(x)\,j +
\sin\vartheta(x)\,k\,.
\end{equation}
That is, we have introduced a fixed basis $i$, $j$ and $k = ij$, 
and found that the remaining degrees of freedom reside in the complex
phase of $j'$, and in the transformed vector potentials,
$\bar{A}$, $B$ and $B'$.  The Lagrangian has become
\[
{\cal L}_{\sc q} = -\frac{1}{4}(\partial_{\mu}\bar{A}_{\nu}
- \partial_{\nu}\bar{A}_{\mu}
+ B_{\mu}B'_{\nu} - B_{\nu}B'_{\mu})^2
+\{\mbox{permutations of $\bar{A}$, $B$ and $B'$}\}
\]
\begin{equation}
+ \frac{\lambda}{2}(B_{\mu}^2 + {B'}_{\mu}^{2})
+ \frac{\lambda'}{2}(\partial_{\mu}\vartheta + \bar{A}_{\mu})^2
+ \frac{\lambda'}{2}(B_{\mu}\sin\vartheta - B'_{\mu}cos\vartheta)^2\,.
\end{equation}
Here, $\lambda$ and $\lambda'$ are {\em a priori}\/ different,
and we propose that $\lambda' \ll \lambda$.  (In this regard, we
might interpret the work of Finkelstein {\em et al.}\/\cite{fjss}
as being in the $\lambda'\!\rightarrow 0$ limit.)  Then in a low
to middle energy system, we might expect $\bar{A}$ production to 
dominate.  For the remainder of this letter, we shall only consider
the case of $\lambda'\sim 0$.  An investigation of the physical content
of the $|D_{\mu}j|^2$ term is presented elsewhere \cite{bj}
Finally, note that $D\phi$ reduces to $\partial\phi$ when $\phi$ is
real.  

We claim that the overwhelming success of conventional, complex
quantum mechanics and quantum field theory is strongly indicative of the
validity of a complex description of local physics (at least, on scales
presently accessible to laboratory physics). Hence, taking the exotic
quaternionic nature to be a high energy effect, we consider ordinary
matter and gauge fields to be $i(x)$ complex. Our new U(1) sector then
constitutes a ``para-charged" region, with our dark matter candidates $B$
and $B^{\prime}$ being the only fields to exhibit a paracharge at low
energies. However, since we expect that all fields will be fully
quaternionic at some large energy scale we introduce an effective theory
in which standard model physics is coupled to our broken quaternionic
sector.

We follow the general discussion of \cite{hof}, which  explores possible
interaction Lagrangian terms in the context of axionic physics.
In our case, $Q$ and $K$ are pure imaginary
quaternion, which leads us to propose an effective interaction Lagrangian 
which couples field densities to $K^2$,
\begin{equation}
{\cal L}_{\rm int} =
\frac{g_{\sc q}}{4\Lambda_{\sc q}^2}\phi^{\dag}\phi\,K^2
+ \frac{g_{\sc q}}{4\Lambda_{\sc q}^3}\bar{\psi}\psi\,K^2
+ \frac{g_{\sc q}}{16\Lambda_{\sc q}^4}\,{\rm tr}F^2\,K^2\,.
\end{equation}
where ${\rm tr}F^2$ is the contribution from the usual U(1) of
electromagnetism.
The exclusion
of terms such as $\bar{\psi}\sigma\psi\cdot K$ is a constraint imposed
by the formalism of our theory.

The scale $\Lambda_{\sc q}$ is a characteristic quantity in QQFT, which we
shall seek bounds on.  Its role is equivalent to that of
$\Lambda_{\sc qcd}$
in $\overline{{\rm MS}}$ renormalisation schemes.  Our theory is only
effective.  Consideration of the underlying fundamental theory is
of paramount interest, but is expected to give equivalent results
to the effective theory below the $\Lambda_{\sc q}$ scale.  It is
natural to expect $\Lambda_{\sc q}$ to be somewhere between the 
electroweak and Planck scales, but we do not rule out the possibility
that this effective description may be inadequate for a description 
of top quark physics.  This would place nontrivial predictions
of our theory
in an experimentally interesting region.  These interaction terms 
are contact terms, describing mutli-particle production above a mass
pole (governed by $\lambda'$) with $\Lambda_{\sc q}^{-1}$ our perturbation
theory parameter.

Our aim, now, is to find physical systems where the coupling of the 
sectors is sufficient to alter the predictions of standard physics.
As our theory is not limited to extreme high energies, we shall 
next examine some astrophysical systems, of increasing scale, 
with the general idea of expecting to find new physics in systems
with sufficient energy to produce excitations of the exotic fields,
and at scales where the quaternionic curvature becomes noticeable.

\section{Applications}
\subsection{$\bar{A}$ production in the Solar interior} 
We consider the possibility of a new luminosity channel for the Sun.
Treating the solar interior as gas of moderately energetic electrons
and photons, we calculate the production of our exotic vector bosons,
using our effective interaction.  We assume that this astrophysical
system is sufficiently cool to lie well below the thresh-hold for 
$B$, $B'$ production, so that $\bar{A}$ production dominates.  A
comparison with the solar (nuclear) energy production rate will allow us 
to place a bound on the scale of our effective theory, $\Lambda_{\sc q}$.

As our effective Lagrangian involves $N\geq 3$ body decays, we do not
expect reabsorption to occur, and hence this may be an efficient
cooling mechanism for a stellar body.  On the other hand, the paraphotons, $\bar{A}$,
can scatter off para-charged particles before escaping the stellar
interior, which process will be suppressed by powers
$\alpha_{\sc q}/\Lambda_{\sc q}^2$, and we shall not consider this 
effect in what follows.  In fact, scattering of this kind provides
the logical detection strategy; exploitation of the paraphotoelectric
effect.

The physical process we shall 
consider is photon conversion to $\bar{A}$ ``para-photons''.  One might 
also consider Primakoff processes involving fermion loops coupling to 
the $\bar{A}$ field; $\bar{A}$ production by nucleons, which requires 
a model for the coupling of 1st generation quarks to the $\bar{A}$ field
within nuclear matter; paraphotonic bremsstrahlung by electrons in
nuclear electrostatic fields; and by $e^+e^-$ annihilation (though
the last named effect is severely suppressed in the solar interior).
In calculations of stellar production of axions \cite{fwy} it is
stated that compton processes contribute approximately 25\% of the
axion luminosity, with Primakoff processes dominating at low energies.
We shall conservatively accept an uncertainty of a factor of 10
underestimation by ignoring these processes.  In higher temperature stars,
compton-type production dominates and our underestimation is expected
to be reduced.

The matrix element for the process $e\gamma\rightarrow e\bar{A}\bar{A}$
is,
\begin{equation}
|{\cal M}_{\bar{A}\bar{A}}|
= \left|\frac{eg_{\sc q}}{\Lambda_{\sc q}^3}
\left(\bar{u}(p')\frac{1}{p\!\!\!/ + k\!\!\!/ -  m}
\varepsilon\!\!\!/_{\gamma}u(p)\right)
\left(q^{\mu}\varepsilon^{\nu}q'_{\mu}\varepsilon'_{\nu}
- q^{\mu}\varepsilon^{\nu}q'_{\nu}\varepsilon'_{\mu}\right)
\,+\,\mbox{cross term}\:\right|\,,
\end{equation}
where $\varepsilon$, $\varepsilon'$ label the polarisation vectors of
the paraphotons (real paraphotons have transverse polarisations only).

We exploit the large differences between the $\Lambda_{\sc q}$, $m_e$
and $\omega_{\gamma}\approx T_{\odot}$ scales to obtain the cross-section,
\begin{equation}
\sigma_{\bar{A}\bar{A}}
= \left(\frac{2\surd{2}}{7\pi}\right)
\left(\frac{\omega_{\gamma}}{m_e}\right)^{5/2}
\alpha\alpha_{\sc q}\frac{m_{e}^4}{\Lambda_{\sc q}^6}\,.
\end{equation}

Calculation of the cooling rate achieved by this dark channel is by
thermal averaging \cite{fwy},
with $E_{\bar{A}\bar{A}} \approx \omega_{\gamma}$ 
representing 100\% efficient photon conversion into paraphotons,
$\dot{\epsilon}_{\bar{A}\bar{A}}(T)
= \frac{1}{\rho}\int\!\!dn_{e} dn_{\gamma}
|v|E_{\bar{A}\bar{A}}\sigma_{\bar{A}\bar{A}}$, so
\begin{equation}
\dot{\epsilon}_{\bar{A}\bar{A}}(T)
=\left(\frac{2\surd{2}}{7\pi}\frac{\zeta(13/2)\Gamma(13/2)}{\pi^2}\right)
\frac{n_e}{\rho}\left(\frac{T_{\odot}}{m_e}\right)^{13/2}
\alpha\alpha_{\sc q}\frac{m_{e}^{8}}{\Lambda_{\sc q}^6}\,,
\end{equation}
where we have used a thermal spectrum for $dn_{\gamma}(T)$, and $|v| = c$.

The nuclear energy production rate in the Sun is 
$\dot{\epsilon}_{\rm nuc}\approx 2\,{\rm ergs}\,
{\rm gm}^{-1}{\rm cm}^{-3}$.  Assuming an effective temperature of
$T_{\odot} = 10^7\,{\rm K}$ 
and using a simple model for the Sun's chemical composition (70\%
Hydrogen, 30\% Helium), we obtain the bound
\begin{equation}
1.81\,10^{20}\,\alpha_{\sc q}\left(m_e/\Lambda_{\sc q}\right)^6 < 2\,,
\end{equation}
which we may convert to a lower bound on $\Lambda_{\sc q}$, assuming 
$\alpha_{\sc q} = {\rm O}(1)$:
$\Lambda_{\sc q} > 2\,{\rm GeV}$.

Note that $\Lambda_{\sc q}$ scales with temperature as
\begin{equation}
(\Lambda_{\sc q}/\Lambda_{\sc q}^{\odot})^6
= (\rho/\rho_{\odot})
(\dot{\epsilon}_{\rm nuc}^{\odot}/\dot{\epsilon}_{\rm nuc})
\left(T/T_{\odot}\right)^{13/2}\,.
\end{equation}
If we consider a red giant star, we have
$T_{\sc rg}\approx 10-100T_{\odot}$,
$\rho_{\sc rg}\approx 100\rho_{\odot}$, and
$\dot{\epsilon}_{\rm nuc}^{\sc rg}
\approx 10\dot{\epsilon}_{\rm nuc}^{\odot}$, 
and hence we have found a better bound:
$\Lambda_{\sc q}^{\sc rg} > 200\,{\rm GeV}$.

We conclude that a $\Lambda_{\sc q}$ of the order of the electro-weak
scale, $\Lambda_{\sc ew} = 250\,{\rm GeV}$, is consistent with
astrophysical bounds.  Futher, there is hope for signatures of new physics
in the current generation of particle accelerator laboratories, in the
form of missing energy not accounted for by neutrinos or neutral hadrons.

\subsection{Cosmological population of exotic species}
In the previous section, we have obtained a lower bound on the scale of 
our effective theory, which indicates that our theory is weakly coupled to
the visible sectors.  Consequently, any relic population of the $\bar{A}$,
$B$, and/or $B'$ bosons would have decoupled early in the thermal 
history of the universe.  Consistency of our model with observational 
cosmology allows us to place bounds on the parameters occurring in
the quaternionic free field Lagrangian Eq.\ (\ref{Lfree}).

Regarding a relic population of almost massless $\bar{A}$ paraphotons,
one can apply results from graviton physics \cite{kt}.  The temperature
of the relativistic species is
$T_{\bar{A}} = (3.91/g_{*s})^{1/3}T_{\gamma}$,
where
$g_{*s} = \sum_{b}g_b(T_b/T_{\gamma})^3
+ (7/8)\sum_{f}g_f(T_f/T_{\gamma})^3$
counts the degrees of freedom in
all relativistic species at the decoupling temperature.
When $T\approx\Lambda_{\sc q} = 200\,{\rm GeV}$, we have 4 quarks,
6 leptons
and all the standard model bosons relativistic, so that $g_{*s} = 90$
and $T_{\bar{A}} \approx 1\,{\rm K}$. 

The contribution of a gas of $\bar{A}$ radiation at this temperature
is insufficient to over-close the universe.  If we allow a small mass,
which in our model corresponds to introducing $\lambda'>0$,
we have the bound 
$m_{\bar{A}} < (g_{*s}/g_{\bar{A}})\,12.8\,{\rm eV} \approx 1\,{\rm keV}$
(see \cite{kt}).
The deeper implications of $\lambda'\neq 0$ are fascinating, as here we
are restoring a dynamical $j$ field, whose broader phenomenological 
implications include the possibility of cosmic strings \cite{bj}.

Regarding the $B$ amd $B'$ fields, we can say that a relic population
of heavy bosons will contribute to the cold dark matter content of the
universe.  We expect broad agreement with heavy neutrino calculations
\cite{kt}; the need to avoid over-closure translates to
a substantial lower bound on $\lambda$ in Eq.(\ref{Lfree}).
A large difference in the $\lambda$ and $\lambda'$ scales may seem
unnatural, and hence undesireable.  This can be remedied by 
considering the case where all of $\bar{A}$, $B$ and $B'$ are light,
hot species, which we will not look at in this paper.

\section{Introduction of a Two Scale Quantum Mechanics}

In addition to the particle phenomenological effects of a broken SU(2),
from which paraphotons and massive spin-1 particles are introduced, we
will further exploit the extended complex structure of quaternions by
exploring the possibility of a two scale quantum theory which is bound to
the broken nature of our phenomenological construction. In this way we
differ from previous quaternionic quantum mechanical approaches by
demanding that, in our broken theory, the general quaternionic quantum
theory is broken into two, orthogonal, complex quantum theories i.e.\ 
commensurate with the symmetry breaking SU(2)$\rightarrow$U(1) the quantum
mechanical Hilbert space {\em locally} breaks to two orthogonal parts.
This is
consistent with our effective construction where fields may be described
as $i$ complex or, as with $B$ and $B^{\prime}$, $j$, $k$ complex.
Denoting by ${\bf C}_{i}$ the complex numbers arising about $i$ and
${\bf C}_{H}$ those about $j$ and $k$, i.e.\ 
$jA_{j} + kA_{k} = k(A_{k}+iA_{j})$, we can
represent our generalised quantum mechanics via the set of numbers
$({\bf C}_{i}, k{\bf C}_{i})$ , where $k{\bf C}_{i} \cong {\bf C}_{H}$,
which can be expressed as two orthogonal contributions via
the anticommutator.

We note that, in our effective theory at least, each field is solely
contained within one or the other quantum region. We therefore can express
the functional dependencies as:
\begin{eqnarray}
{\cal O} \in {\bf C}_{i} \rightarrow {\cal O} = {\cal O} (q_{1}, p_{1})
\;\; : \;\; [q_{1}, p_{1}]=i\hbar_{1} 
\;\; \nonumber \\
{\cal O} \in {\bf C}_{H} \rightarrow {\cal O} = {\cal O} (q_{2}, p_{2})
\;\; : \;\; [q_{2}, p_{2}]=i\hbar_{2} \;\;
\end{eqnarray}

Ideally, the orthogonality of the quantum regions can be employed to
ensure that there is no interference between regions. Pursuing this, and
recalling that $i$ has been singled out by our construction, we can
introduce a quantum bracket of the form
$\{ A_{a}, B_{b} \}_{Q} = -(1/2\hbar)\{[A_{a}, B_{b}], i\}i$,
where $a,b = {\bf C}_{i}$ or ${\bf C}_{H}$ indicating which region the
operators in question correspond to. Such a bracket reduces to the usual
commutation relation for $a=b={\bf C}_{i}$ and provides a generalised
quantum bracket for $a=b={\bf C}_{H}$, where $h$ is 
understood to correspond to $\hbar_1$ and $\hbar_2$ in each case respectively.
Importantly, in the later case the result is an element of ${\bf C}_{i}$.
In this way the ambiguity surrounding which complex number should be
employed in interpreting the quantum bracket is removed. For operators
from different regions this quantum bracket vanishes.

An issue arises, however, when we recall that the quaternionic nature of
operators is hidden in the Lagrangian (or Hamiltonian). Indeed, this
implies that
\begin{equation}
d{\cal O}/dt = -(1/2\hbar)\{[{\cal O}, {\cal H}], i\}i \equiv 0\quad
{\cal O} \in {\bf C}_{H}\,,
\end{equation}
so that under this quantum bracket the operators associated with
${\bf C}_{H}$ are static. This provides a way to introduce a new
background energy density but is however not a very interesting case.
We are not, however, precluded from allowing our quaternionic fields
to carry an explicit time dependence.

An alternative candidate quantum bracket which explicitly employs
the quaternionic elements is not immediately obvious. We are compelled,
then, to instead consider the implicit results of the two quantum regions
which we encapsulate via the commutation relations:
$[q_{1},p_{1}] = i\hbar_{1}$ and $[q_{2},p_{2}]=i\hbar_{2}$.
Rather than represent ${\cal O} \in {\bf C}_{H}$ as
${\cal O} = j{\cal O}_{j} + k{\cal O}_{k} \in k{\bf C}_{i}$
we will separate out the hypercomplex part, $k$, so that
${\cal O} = {\cal O}_{k} + i{\cal O}_{j} = {\cal O}(q_{2}, p_{2})$,
and rely upon the usual quantum bracket
$\{A_{a}, B_{b}\}_{Q} = -(1/2\hbar)[A_{a},B_{b}]$,
where the same conventions are observed as before. Note that this bracket
will again vanish for operators from different regions by virtue of
their dependence on different elements of position and momentum. Further,
this bracket is again always an element of ${\bf C}_{i}$. It follows that
\begin{equation}\label{timeev}
d{\cal O}/dt =
-{i\over \hbar_{1}}[{\cal O}, {\cal H}_{1}]
-{i\over \hbar_{2}}[{\cal O},{\cal H}_{2}]
-{i\over \hbar_{\rm av}}[{\cal O}, {\cal H}_{I}]
-{{i\Delta\hbar^{-1}} \over 2}(J_{2}{\cal O} J_{1} - J_{1}{\cal O}J_{2})
\end{equation}
where $\hbar_{\rm av} = 2\hbar_{1}\hbar_{2}/\hbar_{1}+\hbar_{2}$,
$\Delta\hbar^{-1} = 1/\hbar_{2} - 1/ \hbar_{1}$,
${\cal H}_1$ refers to the
free Hamiltonian in region 1, similarly for $ {\cal H}_2$ and
${\cal H}_{I}$ is the interaction Hamiltonian between the two quantum
regions. $J_{1}$ and $J_{2}$ are the currents from regions one and two
which comprise ${\cal H}_{I}$. It is known that the last term violates
positivity but that unitarity can be restored via the introduction of
noises in the currents. That is, the introduction of noises allows
(\ref{timeev}) to be an element of the class of Lindblad master equations,
allowing embedding into an enlarged unitary dynamics \cite{lind}.

We interpret this enlarged unitary dynamics as the extended dynamical set
which exists beyond the low energy effective theory considered here. This
is consistent with the notion that perturbations of the currents move
to restore the full dynamics of the theory although this cannot be
realised without explicitly enumerating the complete dynamics at high
energies. Nevertheless, the need to embed within a larger dynamical set
is
consistent with our notion of an effective low energy theory.
Interestingly, our low energy, quantum mechanically broken theory is
no longer reversible.

Two interesting cases now present themselves.
Taking $\hbar_{2}\ll\hbar_{1}$
moves the new quantum mechanical sector to be realised as approximately
classical in nature i.e. the quantum effects are suppressed. One possible
application of such a regime is to the baryon asymmetry question. It is
well known that quaternionic field theories are naturally CP violating.
Further, with $\hbar_{2}\ll\hbar_{1}$, it is to be expected that the decay
lifetime of massive $B$ and $B^{\prime}$ particles will be suppressed,
allowing the interaction time to exceed the expansion time in the early
stages of cosmological evolution. We can thus easily satisfy two of the
conditions which lead to baryon asymmetry.
Conversely, taking $\hbar_{2}\gg\hbar_{1}$ will enhance the quantum field
theoretic effects of our new sector. One immediate consequence would be
to the creation of para-particle pairs in strong time varying
gravitational fields. Indeed, an enhanced superluminescence may ensue,
leading to rapid black hole evaporation rates.

Clearly, many interesting phenomena may be tackled in this way.
The essential point, however, is that we can naturally introduce two
different quantum regimes and that we are not compelled, {\em a priori},
to demand that $\hbar_{2}\equiv \hbar_{1}$. Further, we would expect that
such
an approach would allow greater freedom in the interpretation of
$\Lambda_{Q}$, perhaps allowing this scale to be significantly reduced
while retaining phenomenological consistency.

\section{Summary and conclusion}

By starting from a new fundamental basis, we have extended the standard
particle physics description to a broken quaternionic theory whose
consequences are realised at astrophysical and cosmological scales.
We have
demonstrated that such an approach can naturally introduce new exotic
species, satisfy phenomenological bounds and still contain sufficient
additional structural freedom to incorporate additional interesting
phenomena. It is not clear that a simple phenomenological introduction of
an additional SU(2) theory could accommodate such an interpretation, nor
naturally support the new phenomenological constraints. It is intended
that the possibilities surveyed here will be addressed in future work.


\begin{references}

\bibitem{bvn}   G.\ Birkhoff and J.\ von Neumann,
Ann. Math. 37 (1936) 823.

\bibitem{fjss}  D.\ Finkelstein, J.\ M.\ Jauch, S.\ Schiminovich
and D.\ Speiser,
J.\ Math.\ Phys.\ 3 (1962) 207; 4 (1963) 788.

\bibitem{hb}    L.\ P.\ Horwitz and L.\ C.\ Biedenharn,
Ann.\ Phys.\ 157 (1984) 432.

\bibitem{nj}    C.\ G.\ Nash and G.\ C.\ Joshi,
Int.\ J.\ Theor.\ Phys.\ 31 (1992) 965;
J.\ Math.\ Phys.\ 28 (1987) 2883; 28 (1987) 2886.

\bibitem{beh}   B.\ E.\ Hanlon and G.\ C.\ Joshi,
Int.\ J.\ Mod.\ Phys.\ A8 (1993) 3263.

\bibitem{adl}   S.\ L.\ Adler,
Quaternionic quantum mechanics and quantum fields
(Oxford, New York, 1995), and references therein;
Phys.\ Lett.\ B332 (1994) 358.

\bibitem{bja}   S.\ P.\ Brumby, G.\ C.\ Joshi and R.\ Anderson,
Phys.\ Rev.\ A51 (1995) 976.

\bibitem{bj}    S.\ P.\ Brumby and G.\ C.\ Joshi,
to be published in Found.\ Phys.\/.

\bibitem{hof}   S.\ Hoffmann,
Phys.\ Lett.\ B193 (1987) 117.

\bibitem{fwy}   M.\ Fukugita, S.\ Watamura and M.\ Yoshimura,
Phys.\ Rev.\ D26 (1982) 1840.

\bibitem{kt}    W.\ Kolb and M.\ Turner,
The Early Universe (Addison-Wesley, Reading, 1994) Chaps 3 and 5.

\bibitem{lind} L.\ Diosi
Quantum Dynamics with Two Planck Constants and the Semi-classical Limit,
Budapest CRIP preprint, quant-ph/9503023;
G.\ Lindblad,
Comm.\ Math.\ Phys.\ 48 (1976) 1199. 


\end{references}
\end{document}